\documentclass[pra, twocolumn]{revtex4}
\usepackage{amsmath}
\usepackage{amsfonts}
\usepackage{lipsum}
\usepackage{braket}
\usepackage{bm}
\usepackage{color}
\usepackage{graphicx}

\newcommand{\ignore}[1]{}
\newcommand{\beq}{\begin{equation}}
\newcommand{\eeq}{\end{equation}}

\begin{document}
\title{Dynamical phase transition in Floquet optical bistable systems:\\ An approach from finite-size quantum systems}
\author{Tatsuhiko Shirai$^1$}
\email{tatsuhiko.shirai@aoni.waseda.jp}
\author{Synge Todo$^{2,3}$}
\author{Seiji Miyashita$^{2,3,4}$}
\affiliation{%
$^1$Green Computing Systems Research Organization, Waseda University
27, Waseda-cho, Shinjuku-ku, Tokyo, 162-0042 Japan\\
}
\affiliation{%
$^2$Department of Physics, Graduate School of Science,
The University of Tokyo, 7-3-1 Hongo, Bunkyo-Ku, Tokyo 113-0033, Japan\\
}
\affiliation{%
$^3$The Institute for Solid State Physics, The University of Tokyo, 5-1-5 Kashiwanoha, Kashiwa, Chiba 277-8581, Japan\\
}
\affiliation{%
$^4$The Physical Society of Japan, 2-31-22 Yushima, Bunkyo-ku, Tokyo 113-0034, Japan\\
}

\begin{abstract}
We study a dynamical phase transition in optical bistable systems subject to a time-periodic driving field.
The phase transition occurs in the structure of limit cycle as a function of the frequency of the driving field.
In the thermodynamic limit, a single limit cycle is divided into two separated limit cycles at the transition point.
In finite-size systems, however, there is always a single limit cycle due to the quantum tunneling effect.
We use a Floquet dissipative map, which is a time-evolution operator over one period in a dynamics given by a quantum master equation, and discuss the decay rate of relaxation dynamics into the limit cycle based on the dominant eigenvalue of the map.
We found that the decay rate exhibits qualitatively different system-size dependence before and after the phase transition, and it shows a finite-size scaling of spinodal phenomena around the transition point.
The present work provides a systematic way of studying dynamical phase transition observed in time-periodically driven open systems in terms of the Floquet dissipative map.
\end{abstract}

\maketitle
\section{Introduction}\label{intro}
With the progressive advances in quantum technologies, 
it becomes important to understand quantum many-body systems subject to a time-periodic driving and in contact with a dissipative environment~\cite{vorberg2013generalized, hartmann2017asymptotic, gong2018discrete}.
The time evolution of a system coupled to environment can be described by a quantum master equation~\cite{breuer2002theory, carmichael2013statistical}.
The system is usually relaxed to a time-periodic state with a period of the driving due to the dissipation.
The periodic state is an analog of the limit cycle in classical systems.

The dynamical phase transition in structures of the limit cycles appears in bistable systems.
The optical bistability has been found in cavity systems with an external laser field~\cite{gibbs1976differential, felber1976theory, lugiato1984ii, rempe1991optical, gripp1996anharmonicity, rodriguez2017probing}.
As a function of laser intensity, there is a finite interval with bistability of a high transmission state (HTS) and a low transmission state (LTS).
When the intensity of the input laser is time-periodically modulated beyond the bistable regime, the response of the system is qualitatively different depending on the period of the modulation.
For a slow modulation, the system stays in a different stable state depending on whether the laser intensity is increasing or decreasing.
Thus, the system has a large limit cycle including the HTS and the LTS.
On the other hand, for a fast modulation, the limit cycle of the trajectory is kept in either the HTS or the LTS because there is no sufficient time for making the transition between the states.
Thus there exist two separated limit cycles.
The analogous situation in classical systems was discussed in Ising models with a time-periodically oscillating magnetic field~\cite{tome1990dynamic, sides1998kinetic, chakrabarti1999dynamic}.

In the present study, we show that the Floquet dissipative map is useful to characterize the above mentioned phase transition.
The Floquet dissipative map is given by a time-evolution operator of the quantum master equation~\cite{hartmann2017asymptotic},
and it characterizes the limit cycle and the decay rate of the relaxation dynamics.
In the study of finite systems, there is always a single limit cycle, because the quantum tunneling between the HTS and the LTS causes the mixing of them to a hybridized state.
However, the transition point is estimated by looking at the system-size dependence of the decay rate as a function of the period of the driving field.
The decay rate is exponentially small for a fast modulation, while it is finite for a slow modulation.
Around the transition point, the decay rate exhibits the finite-size scaling, which is consistent with the one of classical (conventional) spinodal phenomenon.
Two separated limit cycles appear in the thermodynamic limit, which we study by a mean-field (MF) analysis.

The composition of this paper is given as follows.
Sec.~\ref{sec:FDP} gives a brief review of the Floquet dissipative map.
Sec.~\ref{sec:model} explains the model of cavity systems.
Sec.~\ref{sec:result} provides an analysis of the phase transition in the structure of the limit cycle by using the Floquet dissipative map.
Sec.~\ref{sec:summary} concludes the paper with a future direction.

\section{Floquet dissipative map}\label{sec:FDP}
In this section, we give a brief review of the Floquet dissipative map~\cite{hartmann2017asymptotic} in order to fix the notation.
We consider a system which is subject to a time-periodic driving and in contact with environmental systems.
We denote $\hat{H}(t)$ and $\rho$ the Hamiltonian and the reduced density matrix of the system of interest, respectively.
The dynamics of $\rho$ is assumed to obey the Lindblad equation~\cite{lindblad1976generators},
\begin{equation}
\left\{
\begin{aligned}
\frac{d\rho}{dt}=&{\cal L}(t)\rho=-i [\hat{H}(t), \rho] +\cal{D}\rho, \\
\cal{D}\rho=&\sum_{a=1}^d \gamma_a \left[ 2 \hat{L}_a \rho \hat{L}_a^\dagger - \{\hat{L}_a^\dagger \hat{L}_a, \rho \}\right],
\end{aligned}
\right.
\end{equation}
where $[\cdot, \cdot]$ and $\{\cdot,\cdot\}$ are the commutator and the anti-commutator, respectively.
We set $\hbar=1$.
The Hamiltonian has a period of $T$;
\begin{equation}
\hat{H}(t)=\hat{H}(t+T).
\end{equation}
The dissipator ${\cal D}$ is given by Lindblad operators $\{ \hat{L}_a \}_{a=\{1,\cdots, d\}}$ and the dissipation strength denoted by $\{\gamma_a\}_{a=\{1,\cdots, d\}}$.
Here, we assume that each environmental system labeled by an index $a$ is independent with each other.
The superoperator ${\cal L}(t)$ is referred to as the Liouville operator, and it has the same period as the Hamiltonian, i.e. ${\cal L}(t)={\cal L}(t+T)$.

The dissipative Floquet map ${\cal F}$ is defined by the time-evolution operator over one period,
\begin{equation}
{\cal F}:={\rm T} \exp \left(\int_0^T {\cal L}(t) dt \right),
\end{equation}
where ${\rm T}$ is the time-ordering operator.
Owing to the time periodicity of ${\cal L}(t)$, the state of the system at $t=nT$ $(n\in \mathbb{N}_0)$ is described by
\begin{align}
\rho (nT)=&{\rm T} \exp \left(\int_0^{nT} {\cal L}(t) dt \right)\rho (0),\nonumber\\
=&\left[ {\rm T} \exp \left(\int_0^{T} {\cal L}(t) dt \right) \right]^n \rho (0)= {\cal F}^n \rho (0).
\end{align}
In this way, the dissipative Floquet map describes the stroboscopic dynamics.
We denote eigenvalues and eigenmodes of ${\cal F}$ by $\{ \lambda_m \}$ and $\{\rho_{m} \}$, respectively.
The eigenvalues of ${\cal F}$ are ordered as
\begin{equation}
1=\lambda_0 \geq |\lambda_1| \geq \cdots.
\end{equation}
Note that $\lambda_0=1$, which represents the stationary state of the dissipative Floquet map ${\cal F}$.

\begin{figure}[t]
\includegraphics[width=0.5\textwidth]{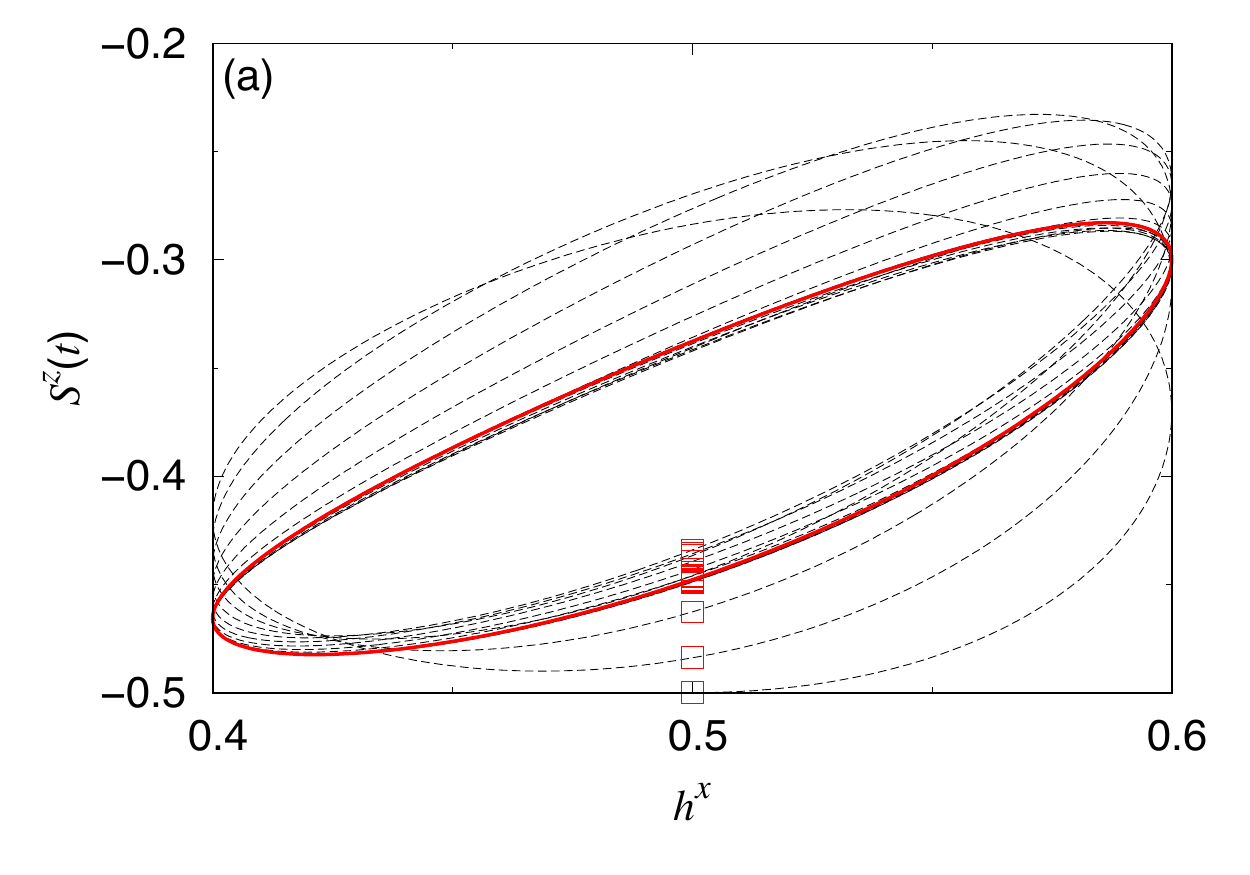}
\includegraphics[width=0.5\textwidth]{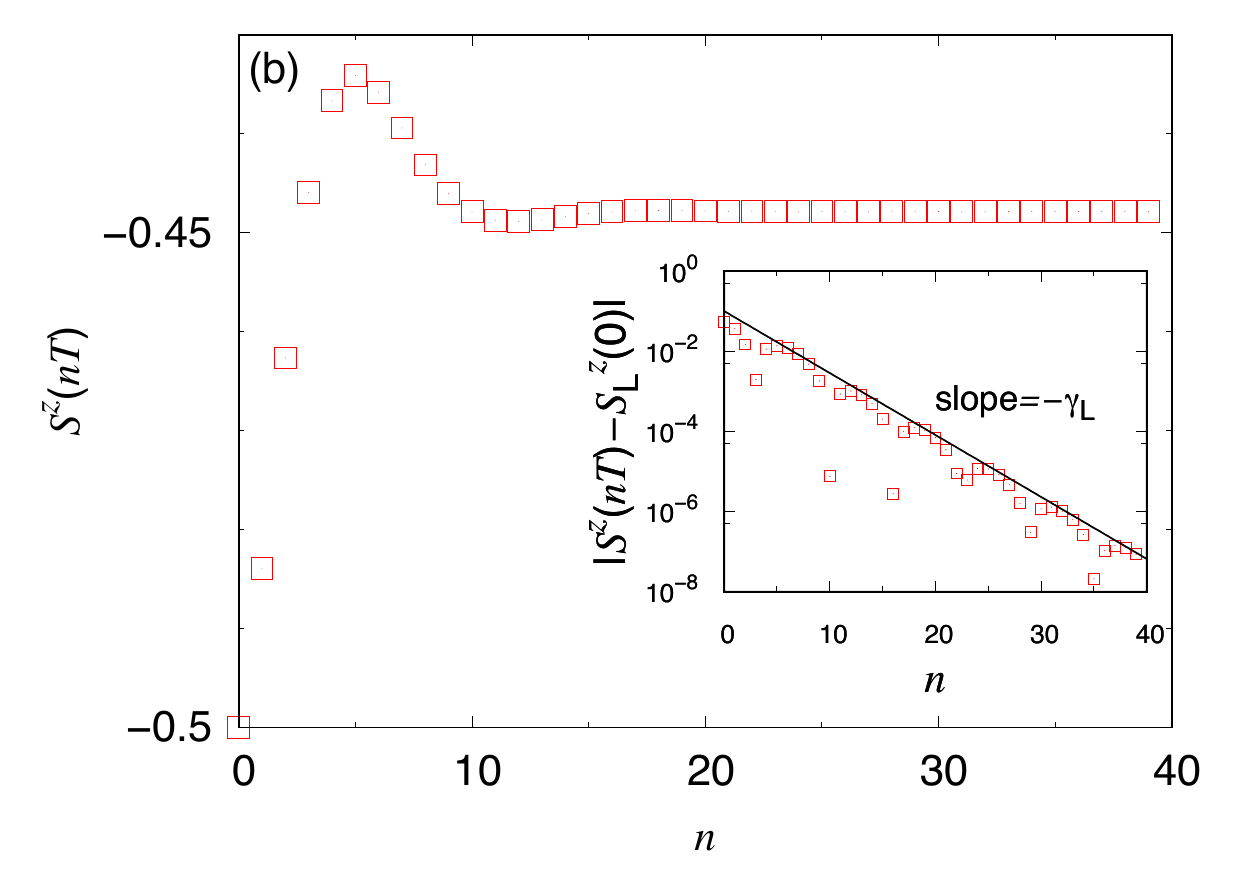}
\caption{
(color online) Demonstration of convergence to a limit cycle:
(a) Time evolution of $S^z(t)$ in the system [Eq.~(\ref{simple})] (dashed line).
The red squares denote the value of $S^z(nT)$ where $n \in \mathbb{N}_0$.
The limit cycle is depicted by red solid line.
(b) The stroboscopic dynamics of $S^z(nT)$.
Inset: The absolute value of the difference between $S^z(nT)$ and $S^z_{\rm L}(0)$ is depicted as a function of $n$.
The black line is a guide to show the damped oscillation with the decay rate $\gamma_{\rm L}$.
}
\label{optical}
\end{figure}

The trajectory of limit cycle is obtained using the eigenmode $\rho_{0}$ as
\begin{equation}
O_{\rm L}(t):=\frac{1}{{\rm Tr}\rho_{0}}{\rm Tr} \left[\hat{O} {\rm T}\exp \left(\int_0^t {\cal L}(\tau) d\tau\right) \rho_{0}\right],\label{limit}
\end{equation}
where $\hat{O}$ is an operator.
Here, $O_{\rm L}(t)$ is the periodic function with period $T$.
The decay rate to the limit cycle is defined by $\lambda_1$ as
\begin{equation}
\gamma_{\rm L} :=-\frac{1}{T}{\rm Re}(\log \lambda_1).\label{decay}
\end{equation}

As a demonstration, we consider a simple generic model, i.e. a $1/2$-spin system with a static magnetic field $h^z$ and time-periodically oscillating magnetic field $h^x(t)$.
The Hamiltonian and the Lindblad operator are given by
\begin{equation}
\left\{
\begin{aligned}
\hat{H}(t)=&h^z \hat{S}^z+ h^x (t) \hat{S}^x,\\
\hat{L}_1=&\hat{S}^-,
\end{aligned}
\right.\label{simple}
\end{equation}
respectively, where
\begin{equation}
h^x(t)=h_0 +h_{\rm ex} \sin \left( 2\pi\frac{t}{T} \right).
\end{equation}
Here, $\vec{S}=\{\hat{S}^x, \hat{S}^y, \hat{S}^z\}$ are the spin-$1/2$ operators and $\hat{S}^- := \hat{S}^x -i \hat{S}^y$.

Figure~\ref{optical}(a) depicts the time evolution of $S^z (t) := {\rm Tr} \hat{S}^z \rho(t)$ for the parameters $(h^z, h_0, h_{\rm ex}, \gamma_1, T)=(1, 0.5, 0.1, 0.05, 6)$.
The initial state is set as the down-spin state, i.e. $\rho(0)=\ket{\downarrow}\bra{\downarrow}$, where $\hat{S}^z \ket{\downarrow}=-1/2 \ket{\downarrow}$.
The system is relaxed to a limit cycle.
The limit cycle is drawn by a red closed trajectory, which is obtained by Eq.~(\ref{limit}).

When the dynamics is observed at stroboscopic times $t=nT$ [red squares in Fig.~\ref{optical}(a)], convergence is clearly observed.
Figure~\ref{optical}(b) shows the approach of $S^z(nT)$ to the limiting value $S^z_{\rm L}(0)$.
It exhibits a damped oscillation with the decay rate $\gamma_L$ [see inset of Fig.~\ref{optical}(b)].
The decay rate obtained by Eq.~(\ref{decay}) is $\gamma_{\rm L} \simeq 0.356 T^{-1}$.

\section{Model}~\label{sec:model}
Next, we study the system exhibiting the optical bistability.
The optical bistablity has been observed in a cavity system in the presence of an external driving field.
The Hamiltonian of the cavity system is divided into a static part and a driven part,
\begin{equation}
\hat{H}(t)=\hat{H}_0+\hat{H}_{\rm ex}(t).
\end{equation}
For the static part, we adopt the Dicke model~\cite{dicke1954coherence}, which was introduced to describe the coupling between an ensemble of $N$ atoms and a single cavity mode,
\begin{equation}
\hat{H}_0=\omega_{\rm ph} \hat{a}^{\dagger} \hat{a} +\omega_{\rm a} \sum_{i=1}^N \hat{S}_i^z +g(\hat{a}^{\dagger}-\hat{a}) \sum_{i=1}^N(\hat{S}_i^+ -\hat{S}_i^-).
\label{Dicke}
\end{equation}
Here, $\omega_{\rm ph}$ is the resonance frequency of cavity mode and $\hat{a}$ and $\hat{a}^\dagger$ are the annihilation and creation operators of bosons, respectively.
Each atom is regarded as a two-level atom, and described using the $1/2$-spin operators $\vec{S}_i=\{\hat{S}_i^x, \hat{S}_i^y, \hat{S}_i^z\}$ and $\hat{S}_i^\pm :=\hat{S}_i^x\pm i\hat{S}_i^y$.
The up-spin state and down-spin state correspond to the excited state and the ground state of the atom, respectively.
The energy gap between the two states is denoted by $\omega_{\rm a}$.
The interaction between photons and atoms is given by the third term in Eq.~(\ref{Dicke}).
The coefficient $g$ is the strength of the interaction.
For the driven part of the Hamiltonian, we adopt the following form~\cite{hassan1978dispersive}:
\begin{equation}
\hat{H}_{\rm ex}(t)=i \xi(t) (\hat{a}^\dagger e^{-i\omega_{\rm ex} t}-\hat{a} e^{i\omega_{\rm ex} t}),
\end{equation}
where the cavity mode is pumped by the external driving field with amplitude $\xi(t)$ and frequency $\omega_{\rm ex}$.
In the present work, we suppose that the driving frequency is the same as both the energy of a cavity photon and a two-level atom, namely,
\begin{equation}
\omega:=\omega_{\rm ex}=\omega_{\rm ph}=\omega_{\rm a},
\end{equation}
where we set $\omega$ as a unit of energy.

We adopt the rotating wave approximation (RWA) to simplify the form of the Hamiltonian, $H(t)$.
For this purpose, we introduce a rotating frame defined by a unitary operator,
\begin{equation}
\hat{U}(t)=\exp \left[-i\omega t \left(\sum_{i=1}^N \hat{S}_i^z +\hat{a}^\dagger \hat{a} \right)\right].
\end{equation}
The Hamiltonian in the rotating frame reads
\begin{align}
\hat{H}_{\rm R}(t)=&\hat{U}^\dagger(t) \left( \hat{H}(t)-i\frac{\partial}{\partial t} \right) \hat{U}(t),\nonumber\\
=&-g \sum_{i=1}^N(\hat{a}^{\dagger} \hat{S}_i^- +\hat{a} \hat{S}_i^+)+ i \xi(t) (\hat{a}^{\dagger}-\hat{a})\nonumber\\
&+g \sum_{i=1}^N(\hat{a}^{\dagger} \hat{S}_i^+ e^{2 i \omega t}+\hat{a} \hat{S}_i^- e^{-2 i \omega t}).
\label{H_R}
\end{align}
In the RWA, we drop the last term in Eq.~(\ref{H_R}).
The RWA is valid when the energy scale of the driving field is much larger than other energy scales, $\omega \gg (g, \xi(t), \dot{\xi}(t)/\xi(t))$.
This condition is not satisfied in the ultra-strong coupling regime, $g \sim \omega$~\cite{ciuti2005quantum}, and/or strong driving field, $\xi \sim \omega$~\cite{shirai2013novel}, but it gives a qualitatively correct description in the parameter regime for the optical bistability.
Then, the Hamiltonian in the rotating frame is given by
\begin{equation}
\hat{H}_{\rm R}(t)=- g \sum_{i=1}^N(\hat{a}^{\dagger} \hat{S}_i^- + \hat{a} \hat{S}_i^+)+ i\xi(t) (\hat{a}^{\dagger}- \hat{a}).
\end{equation}

In the cavity system, the main sources of dissipation are (i) spontaneous emission of each atom from the excited state to the ground state and (ii) loss of photons from the cavity mode.
The dynamics of the dissipative system can be modeled by a quantum master equation in the Lindblad form~\cite{lindblad1976generators, breuer2002theory},
\begin{align}
&\frac{d\rho}{dt}=-i[\hat{H}_{\rm R}(t), \rho] +{\cal D}_{\rm a} \rho+{\cal D}_{\rm ph} \rho,\\
&\left\{
\begin{aligned}
{\cal D}_{\rm a} \rho &:=\gamma \sum_{i=1}^N \left( 2 \hat{S}_i^- \rho \hat{S}_i^+ - \{ \hat{S}_i^+ \hat{S}_i^-, \rho \} \right),\\
{\cal D}_{\rm ph} \rho&:=\kappa \left( 2 \hat{a} \rho \hat{a}^\dagger - \{ \hat{a}^\dagger \hat{a}, \rho \} \right),
\end{aligned}
\right.
\end{align}
where $\rho$ is the density matrix of the cavity system.
The dissipation effects (i) and (ii) are described by the dissipators ${\cal D}_{\rm a}$ and ${\cal D}_{\rm ph}$, respectively.
The dissipation strengths are denoted by $\gamma$ and $\kappa$, respectively.

The system has dynamical variables of photons and spins.
We may study the system directly.
However, in the present work, we adopt the so-called adiabatic approximation to study systems with large number of atoms.
In the adiabatic approximation, we eliminate the degrees of the freedom of photons,
and obtain the Lindblad equation consisting of only atoms~\cite{sarkar1987optical} (see~\cite{cirac1992interaction} for the detailed derivation),
\begin{equation}
\left\{
\begin{aligned}
\frac{d\rho_a}{d(\gamma t)}=&{\cal L}(t) \rho_a :=-i \Omega (t)\sum_{i=1}^N[\hat{S}_i^x, \rho_a] +{\cal D} \rho_a,\\
{\cal D} \rho_a :=&\frac{2C}{N}\sum_{i,j}^N \left( 2\hat{S}_i^- \rho_a \hat{S}_j^+ -\{\hat{S}_i^+ \hat{S}_j^-, \rho_a \} \right)\\
&+\sum_{i=1}^N  \left( 2\hat{S}_i^- \rho_a \hat{S}_i^+ -\{\hat{S}_i^+ \hat{S}_i^-, \rho_a \} \right),\\
\end{aligned}
\right.
\label{bad}
\end{equation}
where $\rho_a := {\rm Tr_{photon}} \rho$.
Here, $C$ and $\Omega(t)$ are referred to as the cavity cooperativity parameter~\cite{carmichael2009statistical} and the Rabi frequency, respectively, and are given by
\begin{equation}
\left\{
\begin{aligned}
C=&\frac{Ng^2}{2\kappa \gamma},\\
\Omega(t)=&\frac{2g\xi(t)}{\kappa\gamma}.\\
\end{aligned}
\right.
\label{C_Omega}
\end{equation}
The adiabatic approximation is valid when the timescale for the photon loss is much faster than other timescales, i.e., $\kappa \gg (\gamma, g, \xi(t))$.
This regime is referred to as the bad-cavity limit of cavity quantum electrodynamics.

In the present study, we solve the dynamics governed by the quantum master equation, Eq.~(\ref{bad}).
To do it numerically, it is necessary to express ${\cal L}(t)$ as a matrix.
Naively, the number of elements in ${\cal L}(t)$ increases exponentially with $N$, which gives a strong restriction on $N$ in the numerical simulation, s.t. $N\lesssim 15$.
However, the present model has a symmetry under exchange of atoms.
It has been known that this symmetry reduces the number of non-zero elements in ${\cal L}(t)$ to the order of $N^3$~\cite{sarkar1987optical,lee2012collective,gegg2016efficient,shirai2018optical}.
In this work, we made use of this property and performed simulations up to $N \simeq 100$.

We study the dynamic response of the optical bistable systems to a time-periodic modulation.
It is known that for the present model the MF analysis gives the exact result in the thermodynamic limit~\cite{mori2013exactness}, i.e. $N\to \infty$ (more precisely, $N\to \infty$ keeping $C$ and $\Omega(t)$ to be constant).
For a system with constant $\Omega$, the optical bistability appears when $C>4$ in the interval $\Omega \in (\Omega_{\rm l}, \Omega_{\rm u})$~\cite{shirai2018optical};
\begin{equation}
\begin{aligned}
\Omega_{\rm l}=&\sqrt{(C^2+10C-2)-C^{1/2}(C-4)^{3/2}},\\
\Omega_{\rm u}=&\sqrt{(C^2+10C-2)+C^{1/2}(C-4)^{3/2}}.
\end{aligned}
\end{equation}
In the present study, we set $C=50$, and thus $(\Omega_{\rm l}, \Omega_{\rm u})=(28.1, 72.1)$.
We consider a sinusoidal modulation of the driving amplitude $\xi(t)$, leading from Eq.~(\ref{C_Omega}) to the expression as
\begin{equation}
\Omega(t)=\Omega_0+\Omega_{\rm ex} \sin \left( 2\pi \frac{t}{T}\right).
\end{equation}
The Rabi frequency oscillates around $\Omega_0$ with the amplitude $\Omega_{\rm ex}$ and the period $T$.
Here, we set $(\Omega_0, \Omega_{\rm ex})=(60, 40)$.
Thus, the oscillation center is within the bistable regime, namely, $\Omega_0 \in (\Omega_{\rm l}, \Omega_{\rm u})$, but the maximum and the minimal values of $\Omega(t)$ are out of the bistable regime; $\Omega_0+\Omega_{\rm ex}>\Omega_{\rm u}$ and $\Omega_0-\Omega_{\rm ex}<\Omega_{\rm l}$.

\section{Result}\label{sec:result}
In this section, using the Floquet dissipative map introduced in Sec.~\ref{sec:FDP}, we study the dynamical phase transition in the cavity system.
This section is divided into three parts.
First, we show the presence of phase transition by looking at the system-size dependence of the decay rate.
Next, we focus on the scaling behavior around the transition point.
Finally, we discuss the phase transition in the thermodynamic limit with MF theory.

\subsection{Phase transition in terms of the decay rate}~\label{sec:A}
\begin{figure}[t]
\includegraphics[width=0.5\textwidth]{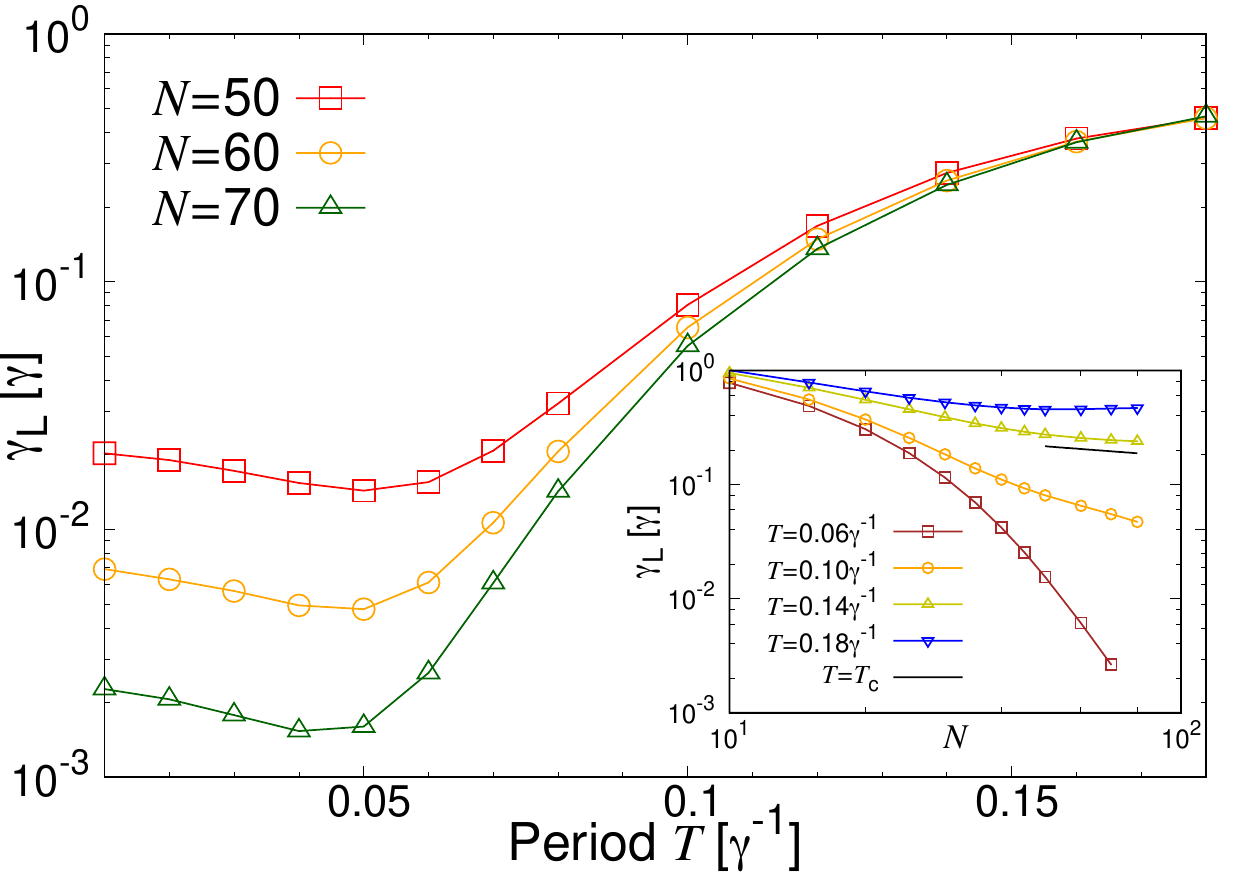}
\caption{
(color online) $T$-dependences of the decay rates $\gamma_{\rm L}$ for different values of $N$.
Inset: the decay rate $\gamma_{\rm L}$ as a function of $N$.
The black solid line denotes the slope estimated by the power-law scaling at the transition point $T=T_{\rm c}$.
}
\label{gap}
\end{figure}

The value of decay rates carries information about phase transitions not only in equilibrium systems~\cite{landau2014guide} but also in non-equilibrium systems~\cite{prosen2008quantum}.
In this subsection, we show that it is also true for the present system.

In Fig.~\ref{gap}, we plot $T$-dependences of the decay rates $\gamma_{\rm L}$ for different values of $N$.
We find a qualitatively different regions for the system-size dependence of $\gamma_{\rm L}$ as a function of the period $T$.
The decay rate becomes exponentially small with the size for small $T$, while it converges to a finite value for large $T$.

The transition point can be estimated by looking at $N$-dependences of the decay rate $\gamma_{\rm L}$ (inset of Fig.~\ref{gap}).
At the transition point $T_{\rm c}$, $\gamma_{\rm L}$ shows a power-law scaling with $N$.
Namely, $\gamma_{\rm L}$ as a function of $N$ changes from concave to convex in the log-log plot.
In the figure, we find that $\gamma_{\rm L}$ is concave at $T=0.06 \gamma^{-1}$, while it is convex at $T=0.18 \gamma^{-1}$.
There exists a transition point between $T=0.06 \gamma^{-1}$ and $T=0.18 \gamma^{-1}$, but it is difficult from our numerical data to give a good estimate of $T_{\rm c}$ due to the limitation of system size.
So, we evaluate the transition point $T_{\rm c}$ from the MF analysis, which gives exact results for the present model in the thermodynamic limit (see the subsection C).
We draw the power-law scaling of $\gamma_{\rm L}$ at the transition point by black solid line in the figure (see the following subsection B).

\begin{figure}[t]
\includegraphics[width=0.5\textwidth]{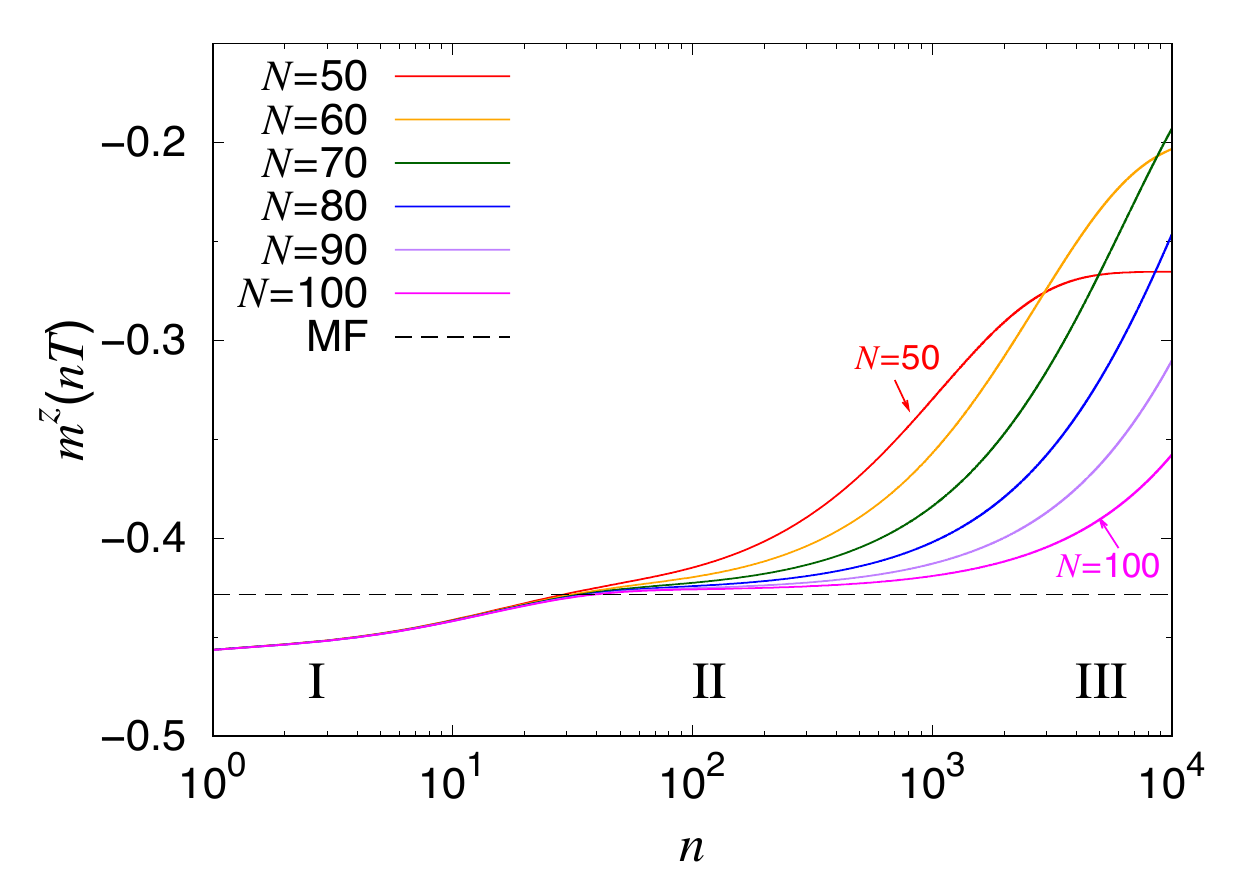}
\caption{
(color online) Stroboscopic dynamics for various values of $N$ at $T=0.06 \gamma^{-1}$.
The timescale of the system being in a metastable state is longer with increasing $N$ from $N=50$ to $N=100$.
The horizontal dashed line gives the value of one of the limit cycles in the MF analysis.
}
\label{timeevo}
\end{figure}

\begin{figure}[t]
\includegraphics[width=0.5\textwidth]{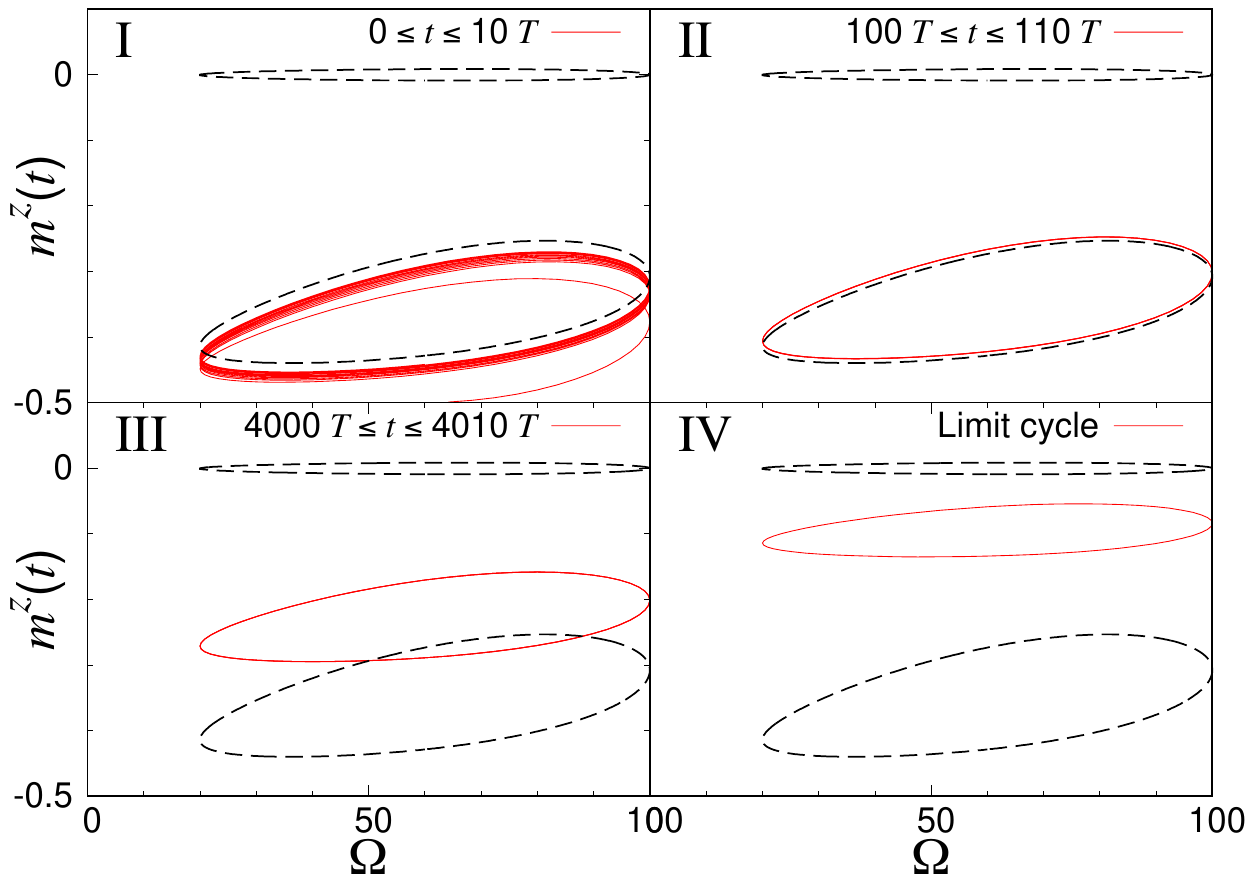}
\caption{
(color online) Trajectories of $m^z(t)$ at $N=70$ and $T=0.06\gamma^{-1}$ in each time domain I, I\hspace{-.1em}I, I\hspace{-.1em}I\hspace{-.1em}I, and I\hspace{-.1em}V (Limit cycle):  I. $0\leq t \leq 10T$, I\hspace{-.1em}I. $100T \leq t \leq 110T$, I\hspace{-.1em}I\hspace{-.1em}I. $4000T \leq t \leq 4010T$.
The limit cycles in MF analysis are shown by dashed lines.
}
\label{trajectory}
\end{figure}

The small decay rate for $T<T_{\rm c}$ implies the existence of a long-lived metastable state.
For example, at $T=0.06 \gamma^{-1}$ and $N=70$, the decay rate is $\gamma_{\rm L}\simeq 3\times10^{-3} \gamma$, i.e. the relaxation time is estimated as $\gamma_{\rm L}^{-1}=3\times10^2 \gamma^{-1} \simeq 5\times10^3 T$.
In order to show this extremely slow relaxation, we calculate the time evolution of spin expectation values,
\begin{equation}
\vec{m}(t)=(m^x(t), m^y(t), m^z(t)) := \frac{1}{N}\sum_{i=1}^N {\rm Tr}\vec{S}_i \rho_a(t).
\end{equation}
We set the initial state as a spin-down state, i.e. $(m^x(0), m^y(0), m^z(0))=(0, 0, -0.5)$.
Figure~\ref{timeevo} depicts $m^z(nT)$ as a function of $n$, which gives the stroboscopic dynamics of $m^z(t)$, for various system sizes at $T=0.06 \gamma^{-1}$.
The horizontal dashed line gives the value of $m^z(nT)$ of one of the limit cycles in the MF analysis.
We find three time regimes denoted by I, I\hspace{-.1em}I, and I\hspace{-.1em}I\hspace{-.1em}I.
In the first time regime I, the system is relaxed to a metastable state.
Here, there is no system-size dependence.
In the second time regime I\hspace{-.1em}I, the value of $m^z(nT)$ is almost unchanged, which implies that the system keeps staying in the metastable state.
The lifetime of the metastable state increases with $N$.
The metastable state becomes one of the limit cycles in the limit $N\to\infty$ (i.e. in the MF analysis).
The MF dynamics will be given explicitly in the next figure.
In the third time regime I\hspace{-.1em}I\hspace{-.1em}I, the system escapes from the metastable state.
It it noted that the figure is depicted in log-scale on $x$-axis, and thus the escape rate is extremely low.
The escape rate is given by the decay rate shown in Fig.~\ref{gap}.

In Figs.~\ref{trajectory}, we plot by red solid lines the trajectories of $m^z(t)$ at $N=70$ and $T=0.06\gamma^{-1}$ in each time regime, I, I\hspace{-.1em}I, I\hspace{-.1em}I\hspace{-.1em}I, and the limit cycle I\hspace{-.1em}V.
In each figure, we plot the trajectories of limit cycles in MF analysis (MF limit cycles) by black dashed lines.
In the first time regime I ($0\leq t \leq 10T$), the trajectory of $m^z(t)$ starting from $m^z(0)=-0.5$ approaches one of the MF limit cycles.
In the second time regime I\hspace{-.1em}I ($100\leq t \leq 110T$), where the system is in the metastable state, the trajectory of $m^z(t)$ is close to one of the MF limit cycles.
In the third time regime I\hspace{-.1em}I\hspace{-.1em}I ($4000\leq t \leq 4010T$), the trajectory is away from the metastable state.
Owing to the extremely small decay rate, the trajectory during $10$ periods seems to be unchanged in figure.
Finally, the system reaches a limit cycle (see I\hspace{-.1em}V in figure).
It is noted that there is a single limit cycle, and it is different from both of the limit cycles in the MF analysis.
This is because the quantum tunneling between the two MF limit cycles causes the mixing of them to a hybridized one.

\subsection{Scaling behavior}
In this subsection, we analyze the scaling behavior of $\gamma_{\rm L}$ around the transition point.
In statistical mechanics, the scaling exponent of the decay rate is called dynamical exponent.
In equilibrium systems, the exponent determines the universality class to which a model belongs.
It is of interest to understand whether the universality class is extended to the present non-equilibrium model.

\begin{figure}[t]
\includegraphics[width=0.5\textwidth]{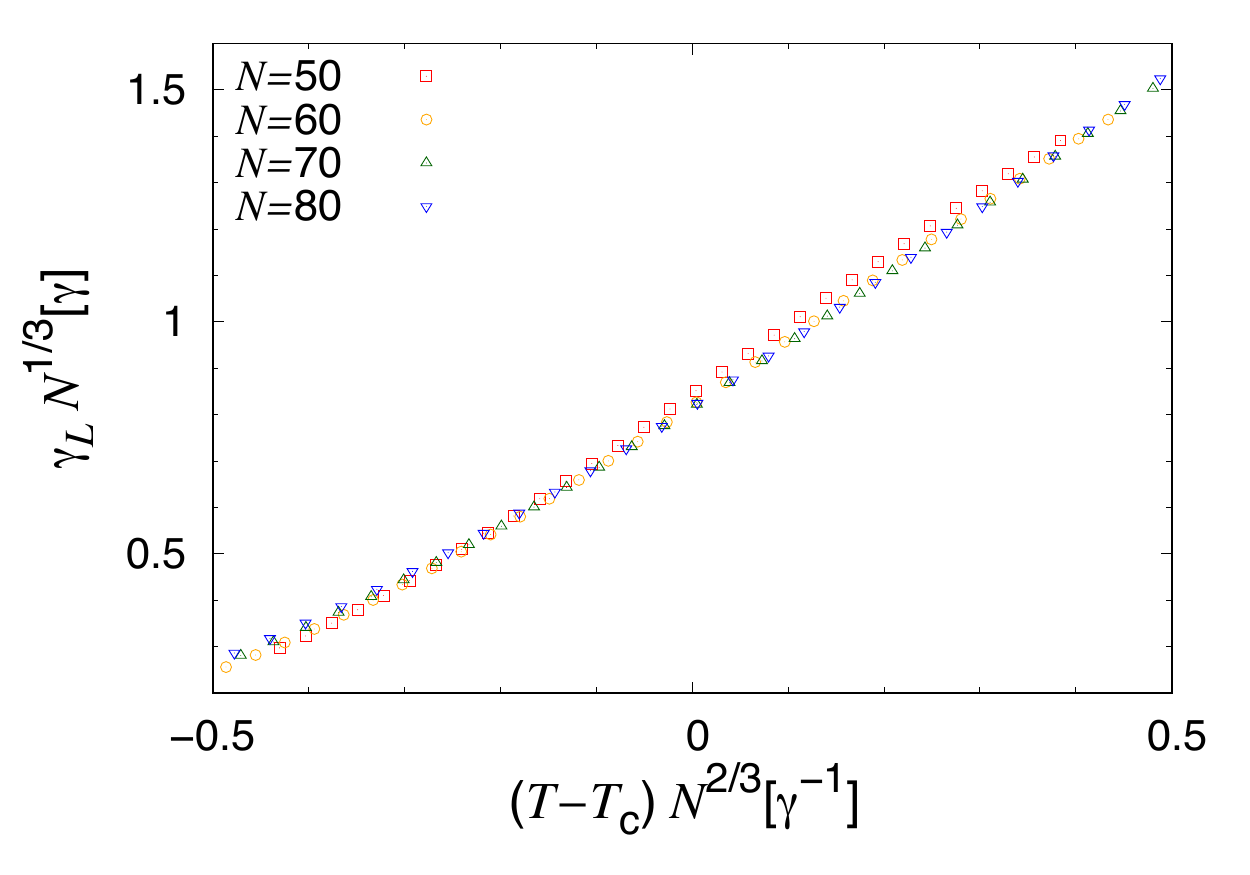}
\caption{
(color online) Scaling plot of $\gamma_{\rm L}$.
All the data collapse well to a scaling function, especially for large values of $N$.
The transition point $T_{\rm c}$ is determined by the MF theory.
}
\label{scaling}
\end{figure}

The argument in the subsection~\ref{sec:A} on the emergence of the metastable state reminds us the spinodal phenomenon.
Thus, we assumed the scaling form of the spinodal phenomenon~\cite{paul1989relaxation, mori2010asymptotic};
\begin{equation}
\gamma_{\rm L} N^{1/3}=f(N^{2/3} (T-T_{\rm c})) \text{ for } T \simeq T_{\rm c},
\end{equation}
where $f(\cdot)$ is a scaling function.
Here, the transition point $T_{\rm c}$ is determined by the MF theory (see the following subsection C).
In Fig.~\ref{scaling}, all the data collapse well to a scaling function, especially for large values of $N$, i.e. the data for $N=70$ and $N=80$ fall on a single line in the scaling plot.
From the scaling form, we obtain
\begin{equation}
\gamma_{\rm L} \sim (T_{\rm c}-T)^{1/2} \text{ as } N\to \infty, \label{scaling_MF}
\end{equation}
and, at the transition point,
\begin{equation}
\gamma_{\rm L} \sim N^{-1/3} \text{ at } T = T_{\rm c},
\end{equation}
which is depicted in the inset of Fig.~\ref{gap} by black solid line.

\subsection{Thermodynamic limit $N \to \infty$}
Finally, we present a MF analysis for the limit cycle and the decay rate of the model, Eq.~(\ref{bad}), corresponding to the thermodynamic limit $N\to \infty$.
In the MF analysis, the density matrix is given by the product state, that is,
\begin{equation}
\rho_a=\otimes_{i=1}^N \rho_{\rm MF},
\end{equation}
where it is assumed that each atom is in the same state.
In the MF approximation, the time evolution is given by
\begin{equation}
\left\{
\begin{aligned}
\frac{dm_{\rm MF}^x}{d(\gamma t)}=&(4C m_{\rm MF}^z-1)m_{\rm MF}^x,\\
\frac{dm_{\rm MF}^y}{d(\gamma t)}=&-\Omega(t) m_{\rm MF}^z+(4C m_{\rm MF}^z-1)m_{\rm MF}^y,\\
\frac{dm_{\rm MF}^z}{d(\gamma t)}=&\Omega(t) m_{\rm MF}^y -4C[(m_{\rm MF}^x)^2+(m_{\rm MF}^y)^2]\\
&-2(m_{\rm MF}^z+1/2),
\end{aligned}
\right.
\label{MF}
\end{equation}
where $\vec{m}_{\rm MF}=(m_{\rm MF}^x, m_{\rm MF}^y, m_{\rm MF}^z) :={\rm Tr} \vec{S}_1 \rho_{\rm MF}$.
After a sufficiently long time, the system is relaxed to a limit cycle with a period $T$
\begin{equation}
\vec{m}_{{\rm MF}, {\rm L}}(t):=\lim_{n\to \infty}\vec{m}_{\rm MF}(t+nT).
\end{equation}

\begin{figure}[t]
\includegraphics[width=0.5\textwidth]{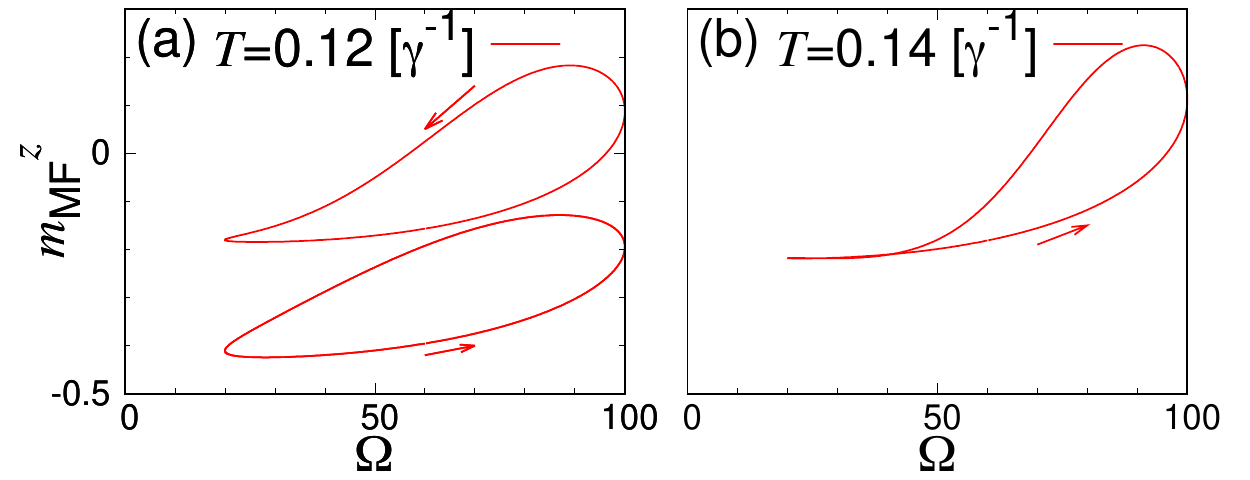}
\caption{
(color online) Trajectories of limit cycles in the MF analysis as a function of $\Omega$ for different values of $T$: (a) $T=0.12 \gamma^{-1} < T_{\rm c}$, and (b) $T=0.14 \gamma^{-1} > T_{\rm c}$.
}
\label{MF_LC}
\end{figure}

\begin{figure}[h]
\includegraphics[width=0.5\textwidth]{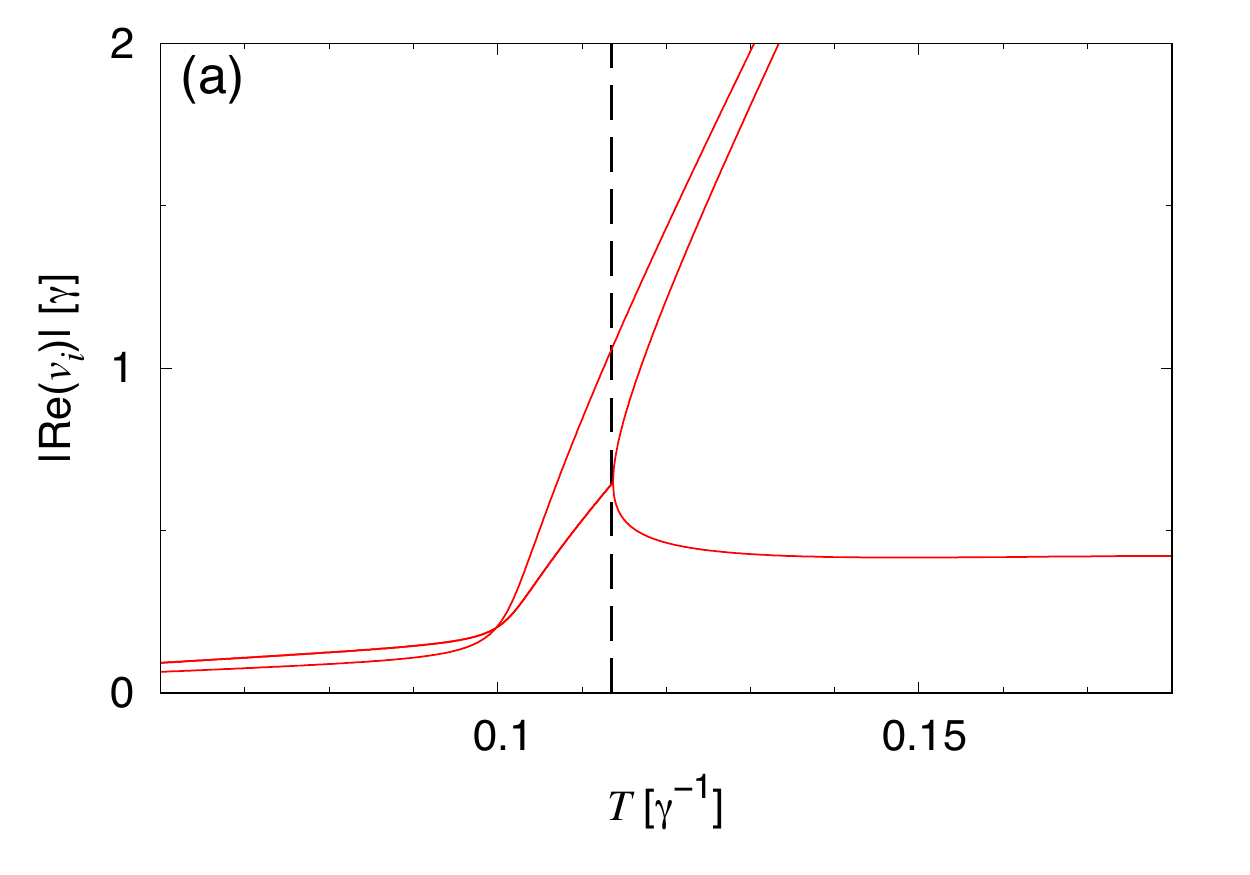}
\includegraphics[width=0.5\textwidth]{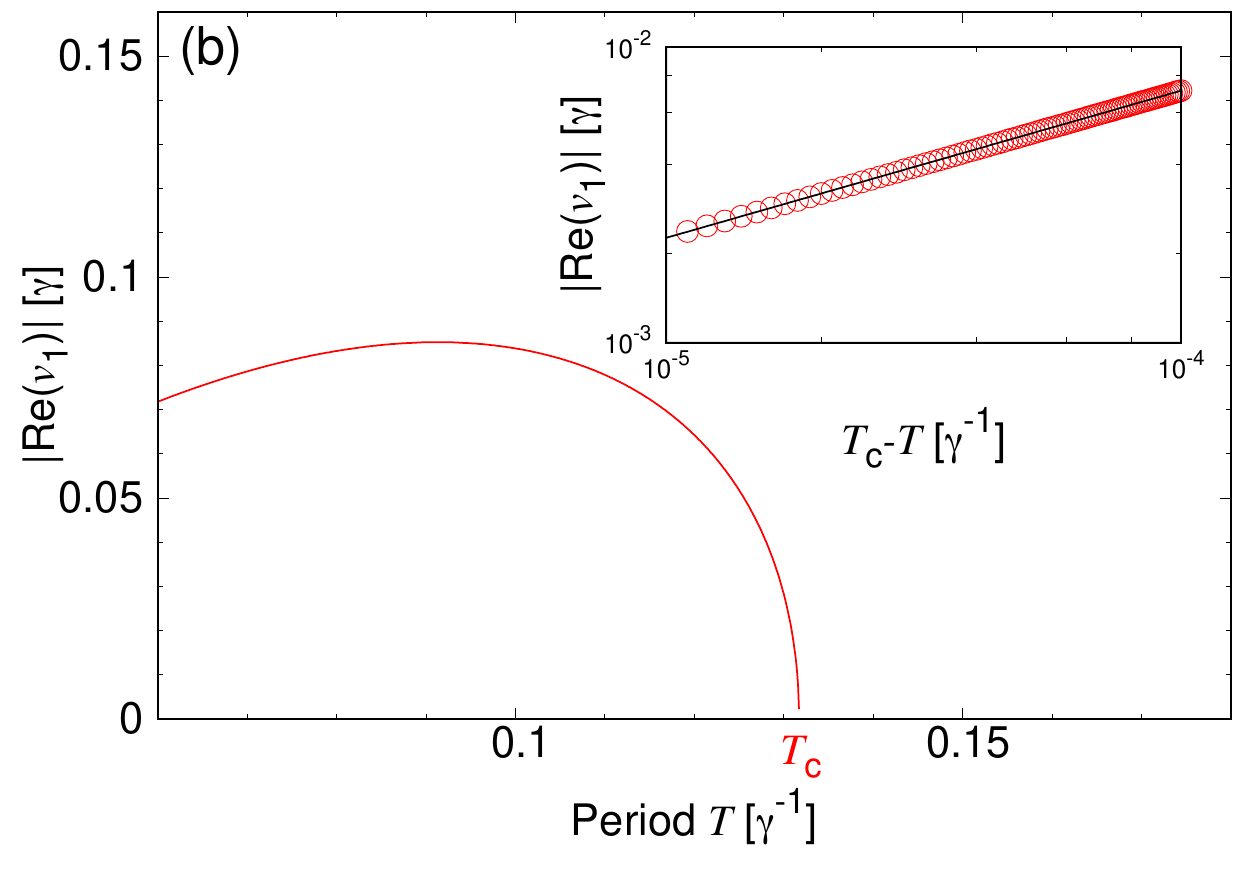}
\caption{
(color online) (a) The decay rates to the limit cycle around $m^z \simeq 0$. The dashed line denotes the value of $T$ where the relaxation dynamics changes from damping oscillation to over damping.
(b) The decay rate to the limit cycle around $m^z \simeq -0.4$, which approaches zero at $T=T_{\rm c}$.
Inset: magnification of data around $T \lesssim T_{\rm c}$. Full curve is a fitting of the dependence $|{\rm Re}(v_1)|$ with a power law, $|{\rm Re}(v_1)| \sim (T_{\rm c}-T)^\alpha$ where $\alpha=0.50$.
}
\label{relaxMF}
\end{figure}

We found a phase transition in the structure of the limit cycle.
For short period ($T=0.12\gamma^{-1}$) [see Fig.~\ref{MF_LC}(a)], the system exhibits two limit cycles.
It is again noted that the separation of the limit cycles appears only in the thermodynamic limit.
On the other hand, for long period ($T=0.14\gamma^{-1}$) [Fig.~\ref{MF_LC}(b)], there is only one limit cycle.

We also perform a linear analysis of the mean-field equation to study the relaxation dynamics to the limit cycle.
First, we apply a perturbation to the limit cycle and set it as an initial state,
\begin{equation}
\vec{m}_{\rm MF}(0)=\vec{m}_{{\rm MF}, {\rm L}} (0) +\delta \vec{e}_{\alpha},
\end{equation}
where $\{ \vec{e}_{\alpha} \}_{\alpha=\{x,y,z\}}$ is a unit vector in $\alpha$-direction and $\delta$ is the perturbation strength.
Then, $\vec{m}_{\rm MF}(t)$ is obtained in the integration of the MF dynamics, Eq.~(\ref{MF}).
The state after one period can be expanded around $\vec{m}_{{\rm MF}, {\rm L}} (0)$ as
\begin{equation}
\vec{m}_{\rm MF}(T)=\vec{m}_{{\rm MF}, {\rm L}} (0)+ \delta T \sum_{\beta=\{x, y, z\}} V_{\alpha\beta} \vec{e}_{\beta} +O((\delta T)^2),
\end{equation}
where $V_{\alpha\beta}$ are the linear response coefficients.
If we regard $\{ V_{\alpha\beta} \}$ as matrix elements of $V$, the matrix $V$ has three eigenvalues denoted by $\{v_i\}_{i=\{1,2,3\}}$.
The real part of each eigenvalue is non-positive and its absolute value describes the decay rate of the corresponding eigenmode.
The decay rates $|{\rm Re} (v_i)|$ are provided for each limit cycle.
Namely, there are six decay rates for a fast driving [e.g. Fig.~\ref{MF_LC}(a)], while there are three decay rates for a slow driving [e.g. Fig.~\ref{MF_LC}(b)].

In Fig.~\ref{relaxMF}, we plot the $T$-dependences of the decay rates $|{\rm Re}(v_i)|$.
Figure~\ref{relaxMF}(a) shows the rate of decay to the limit cycle oscillating around $m^z \simeq 0$.
For $T < 0.114 \gamma^{-1}$, there is a pair of complex conjugate eigenvalues.
But for $T > 0.114 \gamma^{-1}$, all the eigenvalues are real.
Namely, the relaxation dynamics changes from a damping oscillation to over damping with increasing $T$.
Figure~\ref{relaxMF}(b) shows one of the decay rates to the limit cycle around $m^z \simeq -0.4$, which is denoted by $|{\rm Re}({v_1})|$.
In the figure, the other two decay rates are not depicted because they are much larger than $|{\rm Re}({v_1})|$.
The decay rate $|{\rm Re}({v_1})|$ approaches to zero at the transition point $T=T_{\rm c}$.
The estimated value of $T_{\rm c}$ is $T_{\rm c}\simeq 0.13171\gamma^{-1}$.
We also measured the scaling exponent for the decay rate around $T \lesssim T_{\rm c}$;
\begin{equation}
|{\rm Re} (v_1)| \sim (T_{\rm c}-T)^{\alpha},
\label{scalingMF}
\end{equation}
where $\alpha \simeq 0.50$ [see inset of Fig.~\ref{relaxMF}(b)].
This is consistent with the one obtained by the finite-size scaling [see Eq.~(\ref{scaling_MF})].

\section{Summary and Discussion}\label{sec:summary}
We studied the dynamical responses of optical bistable systems to a time-periodic modulation of input driving AC field, and showed a phase transition in the structure of limit cycle as a function of the period of the driving field.
We provided a systematic way of studying the phase transition in terms of the Floquet dissipative map.
We showed that the decay rate, which is given by the dominant eigenvalue of the map, is useful to characterize the phase transition.
The system-size dependence of the decay rate qualitatively changes at the transition point (Fig.~\ref{gap}), and the decay rate exhibits the scaling law of the spinodal phenomenon around the transition point (Fig.~\ref{scaling}).

In the present work, the system was always relaxed to a time-periodic state with the period of the driving field.
However using other set of parameters, at least of the MF level, the system can show different types of long-time asymptotic states such as period doubling and chaos.
The characterization of the phase transition between the periodic state and the non-periodic states in terms of the Floquet dissipative map are directions for future work.

\begin{acknowledgments}
This research was supported by MEXT as the ``Exploratory Challenge on Post-K Computer'' project (Challenge of Basic Science-Exploring Extremes through Multi-Physics and Multi-Scale Simulations), Grants-in-Aid for Scientific Research C (No.18K03444) from MEXT of Japan, and the Elements Strategy Initiative Center for Magnetic Materials (Grant Number 12016013) under the outsourcing project of MEXT.
The authors also thank the Supercomputer Center, the Institute for Solid State Physics, the University of Tokyo, for the use of the facilities.
\end{acknowledgments}

\bibliography{hysteresis.bib}

\end{document}